# Optical atomic coherence at the one-second time scale

Martin M. Boyd, Tanya Zelevinsky, Andrew D. Ludlow, Seth M. Foreman, Sebastian Blatt, Tetsuya Ido, and Jun Ye

JILA, National Institute of Standards and Technology and University of Colorado, and Department of Physics, University of Colorado, Boulder, Colorado 80309-0440

Highest resolution laser spectroscopy has generally been limited to single trapped ion systems due to rapid decoherence which plagues neutral atom ensembles.  Here, precision spectroscopy of ultracold neutral atoms confined in a trapping potential shows superior optical coherence without any deleterious effects from motional degrees of freedom, revealing optical resonance linewidths at the hertz level with an excellent signal to noise ratio. The resonance quality factor of 2.4 x $10^{14}$ is the highest ever recovered in any form of coherent spectroscopy. The spectral resolution permits direct observation of the breaking of nuclear spin degeneracy for the $^1S_0$ and $^3P_0$ optical clock states of $^{87}$Sr under a small magnetic bias field. This optical NMR-like approach allows an accurate measurement of the differential Landé *g*-factor between the two states. The optical atomic coherence demonstrated for collective excitation of a large number of atoms will have a strong impact on quantum measurement and precision frequency metrology.

The relative rates of coherent interaction and decoherence in a quantum system are of fundamental importance for both quantum information science (*1*) and precision metrology (*2*). Enhancing their ratio, which is equivalent to improving spectral resolving power, characterizes much of the recent progress in these fields. Trapped ions have so far provided the best platform for research along this direction, resulting in a number of seminal achievements (*3-8*). The principal advantage of the ion system lies in the clean separation between the internal atomic state and the external center-of-mass motion, leading to long coherence times associated with both degrees of freedom. A large ensemble of neutral atoms offers obvious benefits in the signal size and scalability of a quantum system (*9, 10*). Multi-atom collective effects can also dramatically enhance the coherent matter-field interaction strength (*11*). However, systems based on neutral atoms normally suffer from decoherence due to coupling between their internal and external degrees of freedom (*12*).  In this article we report a record-level spectral resolution in the optical domain based on a doubly forbidden transition in neutral atomic strontium. The atoms are confined in an optical trapping potential engineered for accurate separation between these degrees of freedom (*13*). The large number of quantum absorbers provides a dramatic enhancement in signal size for the recovered hertz-linewidth optical resonance profile.

The demonstrated neutral atom coherence properties will impact a number of research fields, with some initial results reported here. Optical atomic clocks (*14*) benefit directly from the enhanced signal size and the high resonance quality factor. Tests of atomic theory can be performed with increased precision. The available spectral resolution also enables a direct optical manipulation of nuclear spins that are decoupled from the electronic angular momentum. Nuclear spins can have an exceedingly long relaxation time, making them a valuable alternative for quantum information processing and storage (*15*). Two ground state nuclear spins can, for example, be entangled through dipolar interactions when photoassociation channels to high-lying electronic states (such as $^3P_1$) are excited (*16*). Combined with a quantum degenerate gas, the enhanced measurement precision will further strengthen the prospects of using optical lattices to engineer condensed matter systems, e.g. allowing massively parallel quantum measurements.

Much of the recent interest in alkaline earth atoms (and similar atoms and ions such as Yb, Hg, In$^+$, and Al$^+$) arises from the study of the forbidden optical transitions both for metrological applications and as a handle for quantum control, with an important achievement being highly effective narrow line laser cooling (*17-19*). The spin-forbidden $^1S_0$ - $^3P_1$ transition has been extensively studied as a potential optical frequency standard in Mg (*20*), Ca (*21*), and Sr (*22*), and has recently been explored as a tool for high resolution molecular spectroscopy via photoassociation in ultracold Sr (*16*). The doubly forbidden $^1S_0$-$^3P_0$ transition is weakly allowed due to hyperfine-induced state mixing, yielding a linewidth of ~1 mHz for $^{87}$Sr with a nuclear spin of 9/2. This transition is a particularly attractive candidate for optical domain experiments where long coherence times are desirable, and is currently being aggressively pursued for the realization of an optical atomic clock (*23-25*). Furthermore, due to the lack of electronic angular momentum, the level shifts of the two states can be matched with high accuracy in an optical trap (*13*), such that external motions do not decohere the superposition of the two states. Using optically cooled $^{87}$Sr atoms in a zero-differential-Stark-shift 1D optical lattice and a cavity-stabilized probe laser with a sub-Hz spectral width, we have achieved probe-time-limited resonance linewidths of 1.8 Hz at the optical carrier frequency of 4.3x10$^{14}$ Hz. The ratio of these frequencies, corresponding to a resonance quality factor $Q \approx 2.4 \times 10^{14}$, is the highest obtained for any coherent spectral feature.

This ultrahigh spectral resolution allows us to perform experiments in the optical domain analogous to radio-frequency nuclear magnetic resonance (NMR) studies. Under a small magnetic bias field, we make direct observations of the magnetic sublevels associated with the nuclear spin. Furthermore, we have precisely determined the differential Landé g-factor between $^1S_0$ and $^3P_0$ that arises from hyperfine mixing of $^3P_0$ with $^3P_1$ and $^1P_1$. This optical measurement approach uses only a small magnetic bias field, whereas traditional NMR experiments performed on a single state (either $^1S_0$ or $^3P_0$) would need large magnetic fields to induce splitting in the radio frequency range. As the state mixing between $^3P_0$, $^3P_1$, and $^1P_1$ arises from both hyperfine interactions and external fields, the use of a small field permits an accurate, unperturbed measurement of mixing effects. Optical manipulation of nuclear spins shielded by two spin-paired valence electrons, performed with a superior spatial and atomic state selectivity, may provide an attractive choice for quantum information science.

Optical atomic clocks based on neutral atoms benefit directly from a large signal to noise ratio (*S/N*) and a superior line *Q*. Resolving nuclear sublevels with optical spectroscopy permits improved measurements of systematic errors associated with the nuclear spin, such as linear Zeeman shifts, and tensor polarizability that manifests itself as nuclear spin-dependent trap polarization sensitivity. Tensor polarizability of the $^3P_0$ state is one of the important potential systematic uncertainties for fermion-based clocks, and is one of the primary motivations for recent proposals involving electromagnetically induced transparency resonances or DC magnetic field-induced state mixing in bosonic isotopes (*26-28*). The work reported here has permitted control of these systematic effects to ~$5\times10^{-16}$ (*29*). Given the superior *S/N* from the large number of quantum absorbers, we expect this system to be competitive among the best performing clocks in terms of stability. Accuracy is already approaching the level of the best atomic fountain clocks (*30, 31*), and absolute frequency measurement is limited by the Cs-clock-calibrated maser signal available to us via a fiber link (*32*). An all-optical clock comparison is necessary to reveal its greater potential.

To fully exploit the ultranarrow hyperfine-induced transition for high precision spectroscopy it is critical to minimize decoherence from both fundamental and technical origins. The ~100-s coherence time available from the $^{87}$Sr atoms is not yet experimentally practical due to environmental perturbations to the probe laser phase at long time scales, but atomic coherence in the optical domain at 1 s can already greatly improve the current optical clock and quantum measurements. To achieve long atomic coherence times, we trap atoms in an optical lattice with a zero net AC Stark shift between the two clock states, enabling a large number of neutral atoms to be interrogated free of perturbations. The tight atomic confinement enables long probing times and permits spectroscopy free of broadening by atomic motion and photon recoil.

For the highest spectral resolution, it is necessary for the probe laser to have a narrow intrinsic linewidth and a stable center frequency. A cavity-stabilized 698 nm diode laser is used as the optical local oscillator for $^1S_0 - {}^3P_0$ spectroscopy. The linewidth of this oscillator has been characterized by comparison with a second laser operating at 1064 nm (*33*) via an optical frequency comb linking the two distant colors. A heterodyne optical beat signal between the two lasers, measured by the frequency comb, reveals a laser linewidth of < 0.3 Hz (resolution-bandwidth-limited) at 1064 nm for a 3-s integration time. This result demonstrates the ability of the frequency comb to transfer optical phase coherence (~1 rad/s) across hundreds of terahertz. Our frequency comb is also referenced to a hydrogen maser calibrated by the NIST F1 Cs fountain clock (*30*), allowing us to accurately measure the probe laser frequency to $3 \times 10^{-13}$ at 1 s. Additionally, the 698 nm laser has been compared with an independent laser system operating at the same wavelength, revealing resolution-bandwidth-limited laser linewidth of 0.2 Hz, which increases to ~2 Hz for a 30 s integration time. After removing the linear drift, the stability of this local oscillator is about $1 \times 10^{-15}$ from 1 s to 1000 s, limited by the thermal noise of the cavity mirrors (*34*). Thus the probe laser provides the optical coherence needed to perform experiments at the 1 s time scale.

$^{87}$Sr atoms are captured from an atomic beam and cooled to 1 mK using a magneto-optical trap (MOT) acting on the strong $^1S_0 - {}^1P_1$ transition (Fig. 1A). This step is followed by a second-stage MOT using the narrow $^1S_0 - {}^3P_1$ intercombination line that cools the atoms to ~1.5 μK. During narrow line cooling, a nearly vertical one

dimensional lattice is overlapped with the atom cloud for simultaneous cooling and trapping. The lattice is generated by a ~300 mW standing wave with a 60 μm beam waist at the wavelength of 813.428(1) nm, where the $^1S_0$ and $^3P_0$ ac Stark shifts from the trapping field are equal (*35*). The cooling and loading stages take roughly 0.7 s and result in a sample of $10^4$ atoms, spread among ~100 lattice sites. The vacuum-limited lattice lifetime is > 1 s. The atoms are confined in the Lamb-Dicke regime along the axis of the optical lattice. The Lamb-Dicke parameter, or the square root of the ratio of recoil frequency to trap oscillation frequency, is ~0.3. Both the axial and radial trap frequencies are much larger than the $^1S_0$ - $^3P_0$ transition linewidth, leading to the spectral feature comprised of a sharp optical carrier and two sets of resolved motional sidebands. One pair of sidebands is observed ± 40 kHz away from the carrier, corresponding to the axial oscillation frequency in the lattice. The red-detuned sideband is strongly suppressed, indicating that nearly all atoms are in the motional ground state along the lattice axis. The second pair of sidebands at ±125 Hz from the carrier, with nearly equal amplitudes, corresponds to the trap oscillation frequency in the transverse plane.

With atoms confined in the lattice, the linearly polarized (parallel to the lattice polarization) 698 nm laser drives the π transitions (Fig. 1B) for probe times between 0.08 and 1 s, depending on the desired spectral resolution limited by the Fourier transform of the probe time. The effect of the probe laser is detected in two ways. First, after some atoms are excited to the long-lived $^3P_0$ state by the probe laser, the remaining $^1S_0$ population is measured by exciting the strong $^1S_0$-$^1P_1$ transition with a resonant pulse at 461 nm. The $^1S_0$ - $^1P_1$ pulse scatters a large number of signal photons and heats the $^1S_0$

8atoms out of the lattice, leaving only the $^3P_0$ atoms. Once the $^1S_0$ atoms have been removed, the $^3P_0$ population is determined by driving the $^3P_0$ - $^3S_1$ and $^3P_2$ - $^3S_1$ transitions (Fig. 1A) resulting in atomic decay to the ground state via $^3P_1$ for a second measurement using the $^1S_0$ - $^1P_1$ pulse. The second measurement provides superior *S/N* because only atoms initially excited by the 698 nm probe laser contribute to the fluorescence signal, and the zero background is not affected by shot-to-shot atom number fluctuations. Combining both approaches permits signal normalization against atom number fluctuations.

Although the $^3P_0$ and $^1S_0$ states are magnetically insensitive to first order, the hyperfine-induced state mixing, which allows the otherwise forbidden transition, modifies the $^3P_0$ nuclear *g*-factor by about 50%. This results in a linear Zeeman shift in the $^1S_0$ - $^3P_0$ transition of ~ -100 Hz/G per magnetic sublevel $m_F$ (*36, 37*), where we use the convention that the *g*-factor and nuclear magnetic moment carry the same sign and 1 G = $10^{-4}$ T. This effect is shown schematically in Fig. 1B where the 10 nuclear spin sublevels are resolved for the $^1S_0$ and $^3P_0$ states in the presence of a magnetic field. The linear Zeeman shift is an important issue for high resolution spectroscopy as the magnetic sensitivity can cause significant broadening of the transition, as well as line center shifts due to unbalanced population distribution among the sublevels. To achieve the narrowest resonance, the ambient magnetic field must be compensated with three orthogonal sets of Helmholtz coils. An example of this zeroing process is shown in Fig. 2A, where the transition linewidths are measured under various field strengths. After zeroing the field, narrow resonances as in Fig. 2B are routinely obtained. The displayed transition linewidth of 4.5

Hz (full width at half maximum, or FWHM) represents a resonance $Q$ of ~$10^{14}$. The good $S/N$ for the narrow line resonance achieved without any averaging or normalization arises from the contribution of $10^4$ atoms. The ultrahigh spectral resolution has allowed a recent measurement of systematic effects for the optical clock transition at the $9\times10^{-16}$ level (*29*).

The high resolution spectroscopy enables direct measurement of the differential Landé $g$-factor ($\Delta g$) between $^3P_0$ and $^1S_0$. To observe this state mixing effect, a small magnetic field (< 1 G) is applied along the direction of the lattice polarization, and the probe laser polarization is again fixed along this quantization axis to drive $\pi$ transitions. Figure 2C shows a direct observation of the hyperfine-induced state mixing in the form of 10 resolved transition components, with their relative amplitudes influenced by the Clebsch-Gordan coefficients. The narrow linewidth of the forbidden transition allows this NMR-like experiment to be performed optically at small magnetic fields. The magnitude of $\Delta g$ can be measured by mapping out the line splitting vs. magnetic field. Alternatively, eighteen $\sigma^+$ and $\sigma^-$ transitions (Fig. 1B) can be employed to extract both the magnitude and sign (relative to the known $^1S_0$ $g$-factor (*38*)) of $\Delta g$, without accurate calibration of the field. Using the latter approach, we find $\Delta g$ = -108.8(4) Hz/(G $m_F$). The measured $\Delta g$ permits determination of the $^3P_0$ lifetime of 140 (40) s, in agreement with recent *ab initio* calculations (*39, 40*). The uncertainty is largely dominated by inconsistencies among hyperfine mixing models (*36, 37*).

The linewidth of each spectral feature in Fig. 2C is Fourier limited by the 80 ms probe time to ~10 Hz. With the nuclear spin degeneracy removed by a small magnetic field, individual transition components allow exploration of the ultimate limit of our spectral resolution by eliminating any broadening mechanisms due to residual magnetic fields or light shifts, the likely limitation for data such as in Fig. 2B. To reduce the Fourier limit for the linewidth, the spectra of a single resolved sublevel ($m_F = 5/2$ in this case) are probed using π-polarization with the time window extended to ~480 ms. Figures 3 A and B show some sample spectra of the isolated $^1S_0\,(m_F = 5/2) - {}^3P_0\,(m_F = 5/2)$ transition with a Fourier limited linewidth of 1.8 Hz, representing a line $Q$ of ~ $2.4 \times 10^{14}$. This $Q$ is reproduced reliably, as evidenced by the histogram of linewidths measured in the course of one hour (Fig. 3C). Typical linewidths are ~1 – 3 Hz with the statistical scatter owing to residual probe laser noise at the 10-s time scale.

To further explore the limit of coherent atom-light interactions, two-pulse optical Ramsey experiments have also been performed on an isolated π transition. When a system is lifetime limited, the Ramsey technique can achieve higher spectral resolution at the expense of signal-to-noise ratio, leading to useful information on the decoherence process. By performing the experiment in the lattice, the Ramsey interrogation pulse can be prolonged, resulting in a drastically reduced Rabi pedestal width compared to free-space spectroscopy. The reduced number of fringes greatly simplifies identification of the central fringe for applications such as frequency metrology. The fringe period (in Hz) is determined by the sum of the pulse interrogation time, $\tau_R$, and the free-evolution time between pulses, $T_R$, and is given by $1/(\tau_R + T_R)$. Figure 3D shows a sample Ramsey spectrum, where $\tau_R$ = 20 ms and $T_R$ = 25 ms, yielding a fringe pattern with a period of 20.8(3) Hz and fringe FWHM 10.4(2) Hz. For the same transition with $\tau_R$ raised to 80 ms and $T_R$ to 200 ms (Fig. 3D inset) the width of the Rabi pedestal is reduced to ~10 Hz and the recorded fringe linewidth is 1.7(1) Hz.

This linewidth is recovered without significant degradation of the fringe *S/N*, suggesting that the spectral resolution is limited by phase decoherence between light and atoms, and not effects such as trap lifetime. A limit of 1 - 2 Hz is consistent with our measurements of the probe laser noise integrated over the timescales used for spectroscopy. Other potential limitations to the spectral width include Doppler broadening due to relative motion between the lattice and probe beams, and broadening due to tunneling in the lattice. Future measurements will be improved by locking the probe laser to one of the resolved nuclear spin transitions to further suppress residual laser fluctuations. Although the *S/N* associated with a sublevel resonance is reduced compared to a measurement involving all degenerate sublevels, >$10^3$ atoms still contribute to the signal, which allows measurement to proceed without averaging. Improvements to the *S/N* by a factor of ~3 could be achieved by spin-polarizing the atoms into a single sublevel. Employing DC or optical mixing schemes using the bosonic isotope $^{88}$Sr could further enhance the *S/N* by a factor of ~4 due to the increased natural abundance and a non-degenerate ground state.

The line *Q* of ~$2.4 \times 10^{14}$ achieved here provides practical improvements in the fields of precision spectroscopy and quantum measurement. The neutral atom-based spectroscopic system now parallels the best ion systems in terms of fractional resolution but greatly surpasses the latter in signal size.  For optical frequency standards, the high resolution presented here has improved studies of systematic errors for clock accuracy evaluation. With these narrow resonances, clock instability below $10^{-16}$ at 100 s is anticipated in the near future.  For quantum physics and engineering, this system opens the door to using neutral atoms for experiments where long coherence times are necessary, motional and internal atomic quantum states must be controlled independently, and many parallel processors are desired.

41. We thank T. Parker and S. Diddams for providing the NIST hydrogen maser signal. We also thank J. Bergquist, C. Greene, and J. L. Hall for helpful discussions and X. Huang for technical assistance. The work at JILA is supported by ONR, NIST, and NSF. A. D. Ludlow is supported by NSF-IGERT and the University of Colorado Optical Science and Engineering Program. T. Zelevinsky is a National Research Council postdoctoral fellow. T. Ido acknowledges support from Japan Science and Technology Agency. His present address: Space-Time Standards Group, National Institute of Information and Communications Technology (NICT), Koganei, Tokyo.

**Figure Captions:**

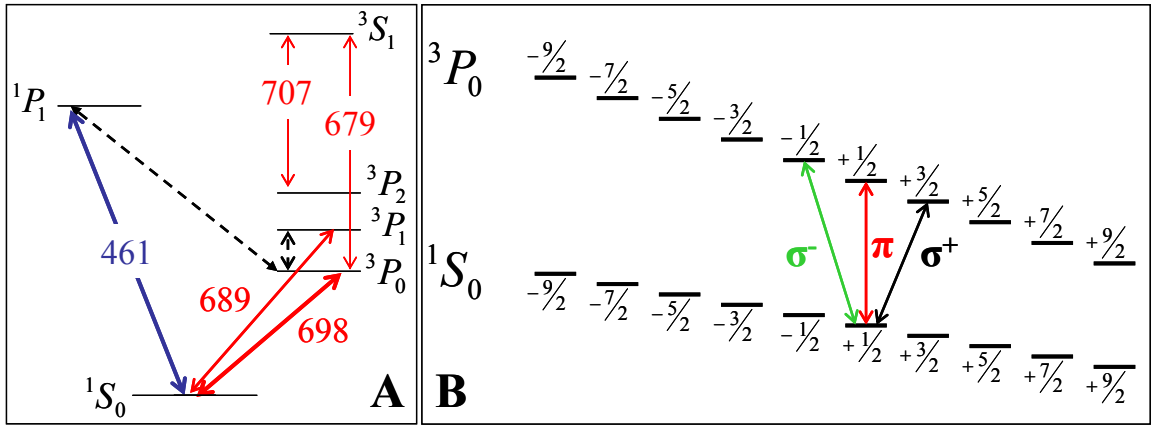

Fig. 1 (A) Partial $^{87}$Sr energy level diagram. Solid arrows show relevant electric dipole transitions with wavelengths in nm. Dashed arrows show the hyperfine interaction-induced state mixing between $^3P_0$ and $^3P_1$ and between $^3P_0$ and $^1P_1$, which provides the non-zero electric dipole moment for the doubly forbidden 698 nm transition. (B) The mixing alters the Landé *g*-factor of the $^3P_0$ state such that it is ~50% larger than that of $^1S_0$, resulting in a linear Zeeman shift for the $^1S_0$-$^3P_0$ transition in the presence of a small magnetic field. The large nuclear spin of $^{87}$Sr (I=9/2) results in 10 sublevels for the $^1S_0$ and $^3P_0$ states, providing 28 possible transitions from the ground state.

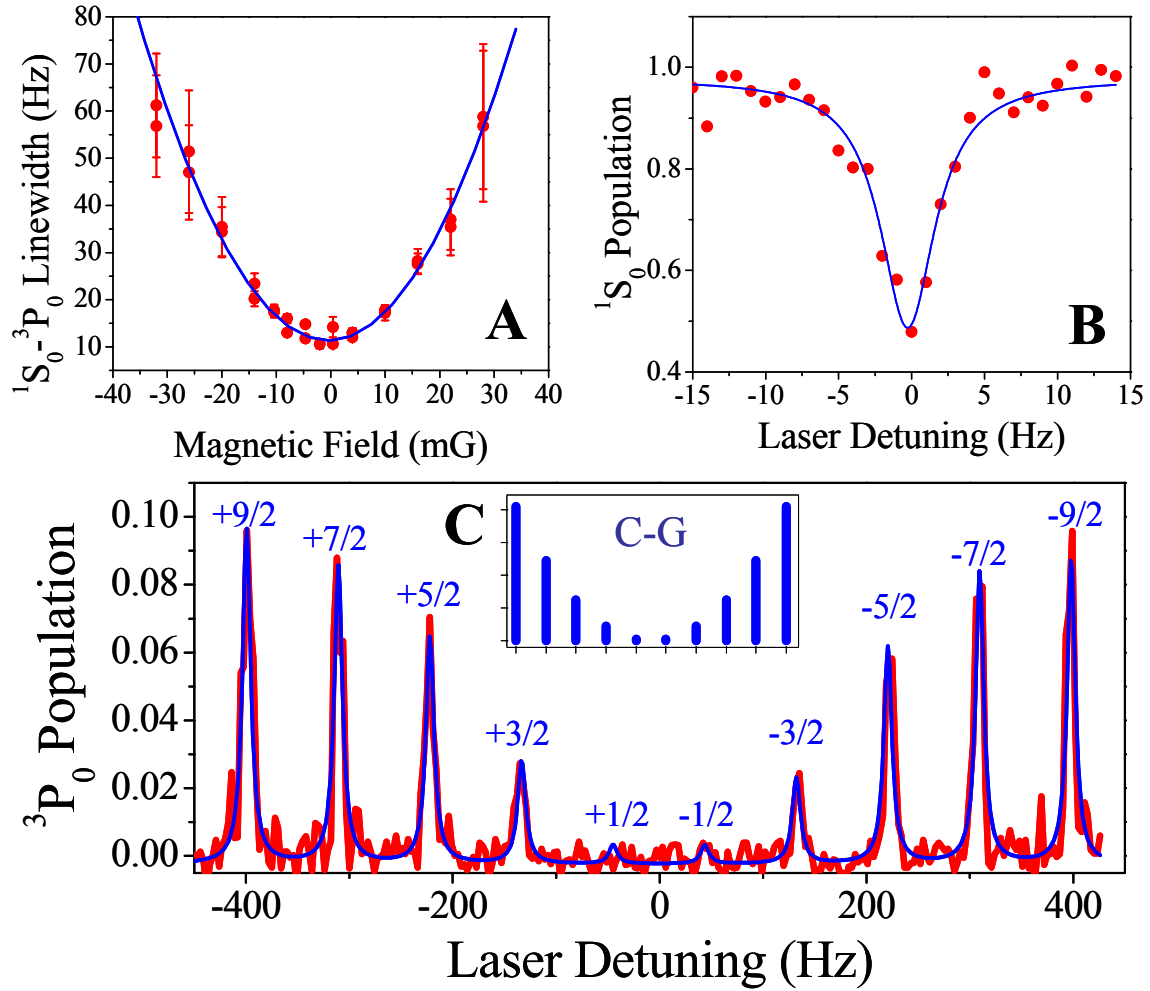

Fig. 2 Spectroscopy of the $^1S_0$-$^3P_0$ transition in $^{87}$Sr. A pair of Helmholtz coils provides a variable field along the lattice (and probe) polarization axis allowing a measurement of the field-dependent transition linewidth as shown in (A) where an 80 ms interrogation pulse is used, limiting the width to ~10 Hz. (B) A representative spectrum when the ambient field is well controlled. Here a longer pulse is used (~480ms, or a 1.8 Hz Fourier limit) but the linewidth is limited to 4.5 Hz by residual magnetic fields and possibly residual Stark shifts. (C) A field of 0.77 G is applied along the polarization axis, and the individual Fourier-limited

(10 Hz) π transitions are easily resolved. Data are shown in red and a fit of 10 evenly spaced transitions is shown in blue. The calculated transition probabilities based on Clebsch-Gordan coefficients are included in the inset. In (B) and (C) the population is scaled by the total number of atoms available for spectroscopy ($\sim 10^4$).

19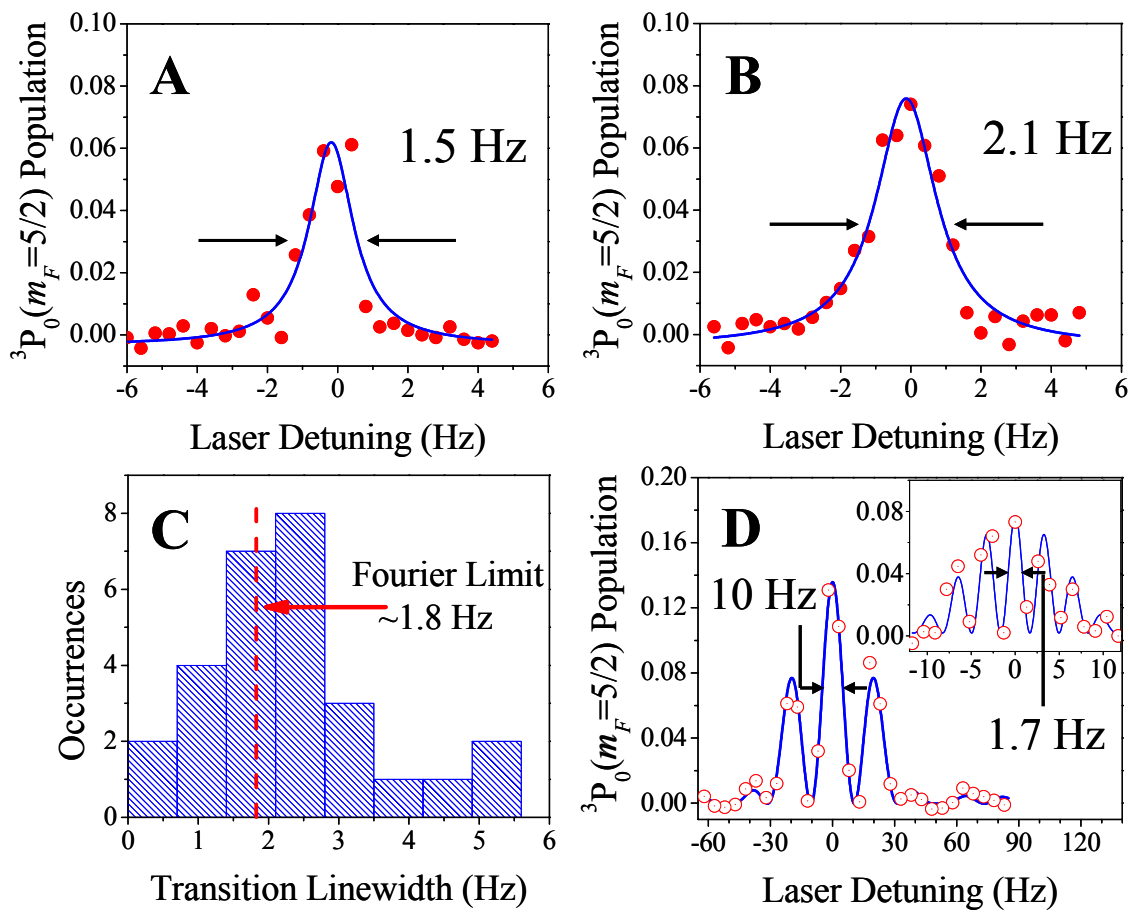

Fig. 3 Spectroscopy of the isolated $^1S_0$ ($m_F = 5/2$) – $^3P_0$ ($m_F = 5/2$) transition. Resolving individual sublevels allows spectroscopy without magnetic or Stark broadening. Spectra in (A) and (B) are taken under identical experimental conditions using a pulse time of 480 ms, and linewidths of 1.5(2) and 2.1(2) Hz are achieved. (C) A histogram of the linewidths of 28 traces within ~1 hour. The average linewidth in (C) is near the 1.8-Hz Fourier limit. (D) Ramsey fringes with a 20.8(3) Hz period and 10.4(2) Hz fringe width, with data shown as open circles. Inset shows a Ramsey pattern with a 1.7(1) Hz fringe FWHM.